# Muscle Cell-Based 'Living Diodes'

*Uryan Isik Can[1], Neerajha Nagarajan[1], Dervis Can Vural[3], and Pinar Zorlutuna\*[1]*

[1]Aerospace and Mechanical Engineering Department, University of Notre Dame, Notre Dame, IN 46556, USA
[2]Department of Physics, University of Notre Dame, Notre Dame, IN 46556, USA

In recent years, there has been tremendous progress in materials science and electronics engineering, in particularly the development of stretchable, flexible electronics that are optimized for interfacing with soft materials.[1, 2] The next step is to expand the domain of such devices to functional biological structures by augmenting, engineering or re-designing "computational tissues", such that part of an organ can act as a programmable arrangement of logic gates or an interface to an external device. Current approaches exploring the use of cell-based elements in creating biocircuits primarily rely on genetic manipulation of the cell, as well as introducing chemicals and other biomolecules to achieve certain functions. Currently, there is no cell-based signal-processing device that operates solely on external electrical triggers. Here, we design, fabricate and characterize a new type of diode that is made entirely of living cells through micropatterned co-cultures of electrically excitable cardiac muscle cells (CMs, i.e. cardiomyocytes) and non-excitable cardiac fibroblasts (CFs). Non-excitable cells can electrically couple to excitable cells through cell-cell junctions and relay signals passively up to a certain distance, though they cannot amplify or propagate external signals directly coming to their membranes. In contrast, excitable cells can initiate, amplify and propagate signals in response to external stimuli through their specialized membrane channels. Our muscle cell-based diode (MCD) design confines these two cell types in a rectangular pattern using a novel, self-forming micropatterning approach, where one side consists of electrically excitable cells and the other side consists of purely non-excitable cells. This configuration allows the signal initiated on the excitable side to pass to the non-excitable side, whereas, it is not possible to pass any signals in the other direction since the cells on the non-excitable side are not able to initiate any action potentials (AP). As such, we find that the controlled arrangement of excitable and non-excitable cell types can be used to transduce electrical signals unidirectionally, essentially achieving a diode function, as we have shown through the characterization of the electrical response of the MCDs using microelectrode arrays (MEA).

Biocomputing is a recent field that emerged around the idea of applying biomolecular systems to information processing.[3-5] Initially limited to chemical reactions that behave as single logic-gates,[6, 7] this technology has moved towards using reaction networks of cell-derived biomacromolecules (such as enzyme complexes) to instantiate multiple logic gates.[8, 9] More recently, genetically modified cells have been used to perform rudimentary computational tasks.[10] Current state-of-the-art biocomputing primarily involves single cells (either bacteria[11-13] or mammalian[14-16]), and information processing is done at the gene or protein level rather than between cell groups. Specifically, information is processed chemically through differential gene and protein expression, in the end controlling the rate of production of certain enzymes.[17]

In addition to chemical signals (which naturally tend to be slow), some cells, such as the CMs, also respond to electrical signals. CMs are excitable cells and can receive an electrical input both internally, through their gap junctions that connect to other cells, and externally, through their voltage-gated ion channels. In contrast, CFs are non-excitable cells and can only receive electrical signals through their gap junctions. Therefore, an electrical signal can be initiated from a CM and transferred to both CMs and CFs. However, such signals cannot be initiated from a CF. The degree of CM – CF coupling is relevant to a number of pathological conditions and has been investigated extensively.[18-25] Although CFs cannot *initiate* an action potential like muscle cells, they can still *propagate* the electrical signal passively, for a limited distance.[18, 21] For cultured rat neonatal ventricular CMs, it was shown that electrical activity could be conducted over a distance of up to 300 μm via CF inserts, causing insert-length dependent delays in wave propagation.[19] This result was reproduced by comparative studies with gap junction deficient cells as controls, confirming the need for gap junctions in electrical conduction.[26]

First, we tested the unidirectional signal transduction potential of specific patterns of these two cell types using a computational model. Our two variable model was based on that of Karma,[24] which can accurately reproduce much of the complex behavior of single excitable cells. We adapted this model to include non-excitable cells and their gap junctions with the excitable cells and simulated the response of an 800 μm long CM – CF chain, where the first 640 μm consisting purely of CMs, and the remaining 160 μm consisting purely of CFs (Figure S1). A signal initiated from the CM end of the chain propagated through the CMs without loss then attenuates through the CFs. When the other end of the chain is stimulated by the same signal, the signal rapidly attenuated. By the time the signal reaches the CM – CF boundary it was orders of magnitude below the CM excitation threshold, and therefore could not propagate through.

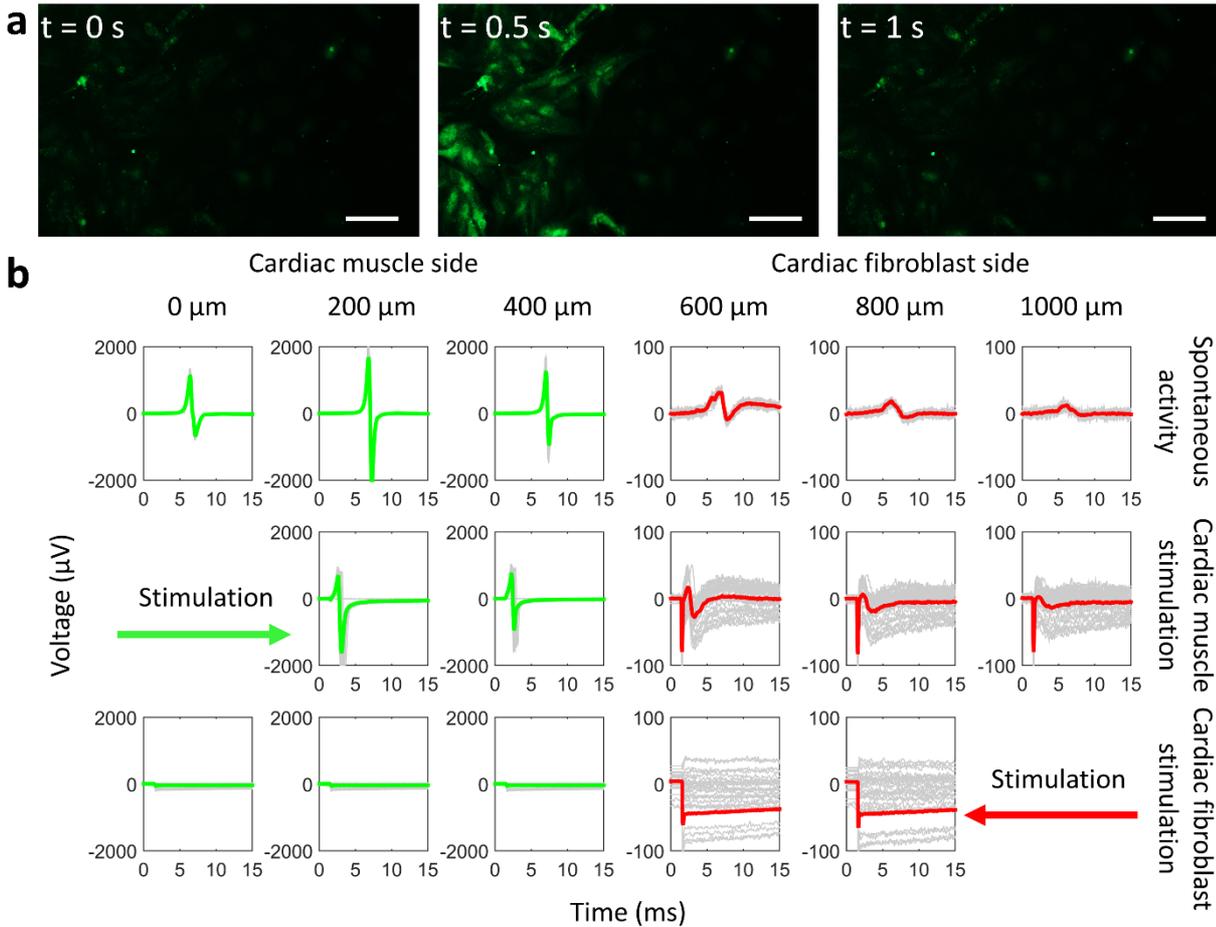

**Figure 1.** $Ca^{2+}$ flux (green) imaging of the micropatterned cardiac muscle cell (CM, left) and cardiac fibroblast (CF, right) co-culture (a, scale bars: 100 μm, see Supplementary Movie 1 for the video file). Baseline activity (b, top), stimulations from excitable CM (b, middle) and non – excitable CF (b, bottom) sides of the co-culture. For all three cases, individual, consecutive AP events (gray) were drawn and averaged (green for CMs or red for CFs).

Second, we investigated the membrane potential transduction in the CM and CF co-culture through both spontaneous and stimulated electrical activity measurements (**Figure 1**). In these experiments, we created two regions containing either excitable or non-excitable cells by partially blocking the MEA substrate surface with a thin poly (dimethylsiloxane) PDMS sheet (Figure S2). To confirm the functionality of the patterned cells, we performed fast $Ca^{2+}$ imaging within the patterned co-culture during their spontaneous beating (Figure 1a, Supplementary Movie 1). This result also confirmed the spatial distribution of excitable and non-excitable cells within the culture.

We measured the extracellular membrane potentials from both the excitable and the non-excitable cells using an MEA which consists of 60 microelectrodes spaced 200 μm apart (Figure 1b). While gray curves represent individual AP events occurring consecutively, the green (measured from excitable side) and the red (measured from non-excitable side) curves are the averages of these signals. Magnitude of the electrical signal decreased upon passing to the CF side and then attenuated over distance, whereas CM side did not show such position dependent attenuation in the signal (Figure 1b, top). Resting potential of CMs and CFs are -60 to -80 mV and -20 to -40 mV, respectively.[21] Therefore, it was expected to see lower extracellular membrane potentials on the CF side than on the CM side. In control studies, we performed similar measurements with samples without any CFs on the non-excitable side and confirmed that the signal read out from the electrode is due to the presence of the CFs, hence due to the signal relayed through the cell-cell junctions, and not an attenuated signal coming from the nearby CMs. To fabricate the controls, we simply kept the PDMS thin film coverage until the day 5 of the culture and removed it just prior to measurements, avoiding any cell presence on the electrodes during the measurements.

We applied electrical stimulations from both excitable (Figure 1b, middle) and non-excitable (Figure 1b, bottom) sides and measured the electrical response and signal prorogation throughout the co-culture. Because of the ability of CMs to exhibit spontaneous electrical activity, we demonstrated signal propagation by stimulating the cells with a higher frequency than their spontaneous electrical activity and used the change in the frequency of the membrane potential to assess the signal propagation. In the forward direction (CM to CF), upon stimulation from the excitable side of the culture, the cardiac cells paced their beating rate to the stimulation frequency. However, in the reverse direction (CF to CM), they did not show such response upon stimulation. This data demonstrates that the CFs were able to passively relay the signals coming through gap junctions whereas they were not able to propagate any signals they received directly as external stimulation. This result agrees with both our computer simulations showing unidirectional signal propagation and the literature on differential excitability of cardiac cells.[27-30]

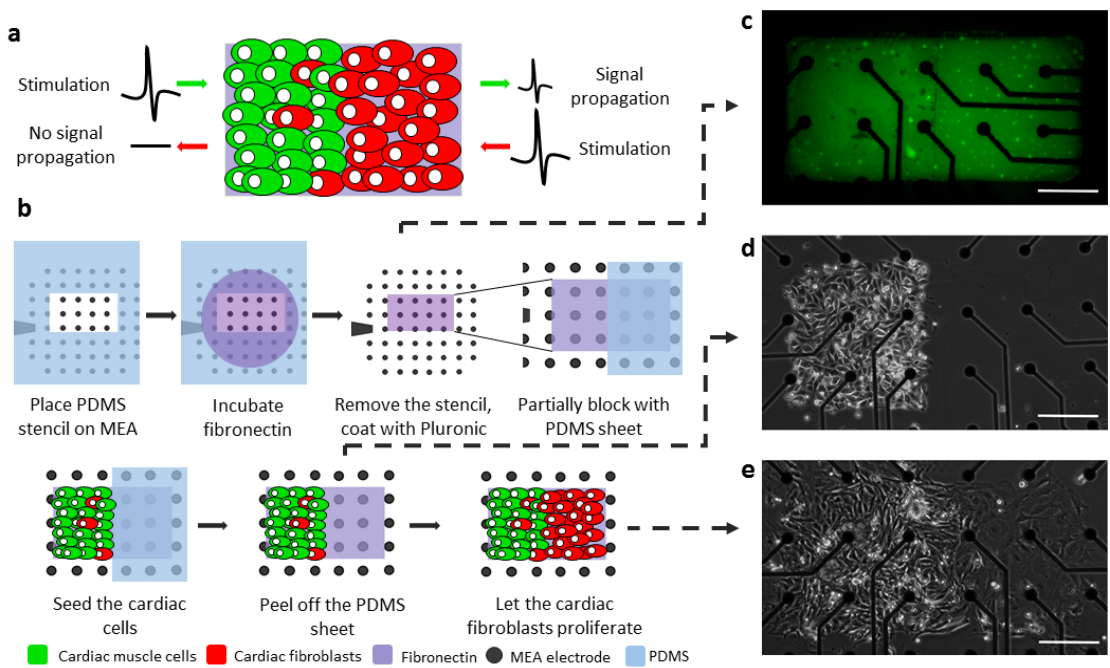

**Figure 2.** MCD design and working principle showing the unidirectional signal propagation (a). Schematic of the co-culture patterning approach to create MCDs on MEAs (b). Fibronectin pattern on the MEA substrate visualized using Alexa-488 tagged fibrinogen (c). Cardiac muscle cell (CM) enriched, cardiac fibroblast (CF) containing cell pattern after the removal of PDMS sheet covering half of the pattern (Day 1, d). Completed MCD structure consisting of CMs and CFs (Day 6, e). (Scale bars: 200 μm).

Next, we designed a modular circuit component, the MCD, where electrically excitable CMs and non-excitable CFs are confined in rectangular micropatterns (**Figure 2**a). Achieving such confined pattern is necessary to isolate this circuit component from signals coming from elsewhere to minimize the error and noise. In order to achieve the isolated components, the first step is to precisely control the distribution of CMs and CFs. However, this is a very challenging task. Current co-culture patterning approaches either confine only one cell type or use sophisticated automated printing methods.[31-33] Furthermore, these methods require a second cell seeding procedure, which causes stress for the first seeded cells and potential cross contamination (one cell type attaching on the other). In our MCD design it is crucial to avoid the presence of CMs in the CF side, since they would render the non-excitable region excitable.

To generate these defined co-cultures of CMs and CFs in rectangular patterns of 500 x 1000 μm we used stencil based protein patterning[34] and partial covering of the protein pattern temporarily[19] in combination with our self-forming micropatterning approach (Figure 2b). Specifically, substrate surfaces were selectively functionalized by fibronectin adsorption for preferential cell attachment using a micropatterned PDMS stencil having 500 x 1000 μm rectangular openings (Figure 2c). To minimize cell attachment and/or growth outside the protein pattern, the substrate surface was treated with an anti-fouling agent (Pluronic F127), and the media was depleted of residual fibronectin prior to cell seeding. A PDMS sheet was then used to partially block the fibronectin pattern in order to populate these micropatterned surfaces with the two different cell types in a controlled manner. After the seeding of the cardiac cell suspension containing 19% ± 1 CFs and 81% ± 1 CMs (n = 3, Figure S3), the PDMS was removed (Figure 2d). In addition to differential excitability of CMs and CFs, these two cell types are also different in terms of their proliferative behavior. Unlike CMs, CFs are highly proliferative. Therefore, cells proliferating across the pattern (Figure 2e) are expected to be only CFs resulting in a purely non-excitable cell population on one end of the MCD. This self-forming patterning approach ensures that there are no excitable cells on the non-excitable end.

Once the MCD was obtained through CF proliferation, we performed double immunostaining on Day 6 (Figure 3a) to examine the distributions of the micropatterned cell populations. Figure 4c shows

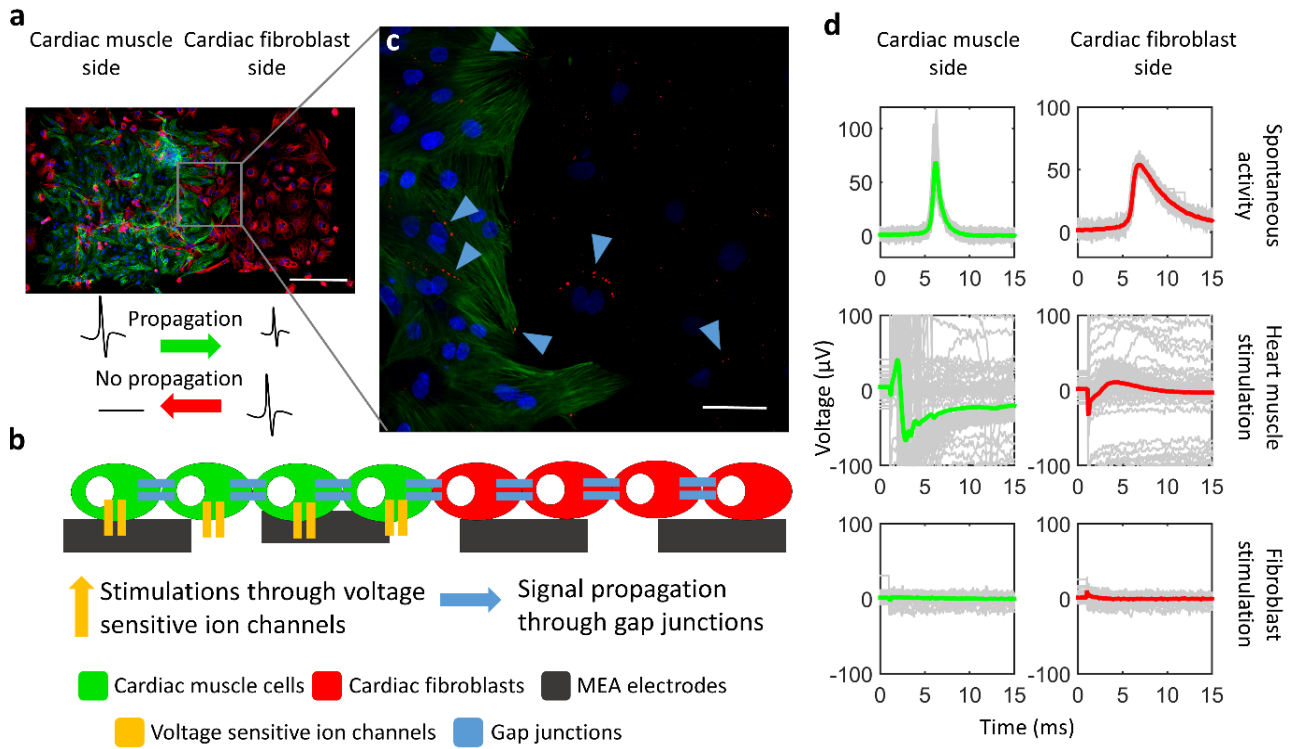

**Figure 3.** Fluorescence image of Troponin-I (green) and Vimentin (red) immunostaining of the MCD counter stained for the cell nuclei (blue) (a, scale bar: 200 μm). Working mechanism of the MCD (b). Fluorescence image of Troponin-I (green) and Connexin 43 (red) immunostaining of the MCD counter stained for the cell nuclei (blue) (c, scale bar: 100 μm). Electrical activity of the MCD (d) measured from cardiac muscle cell (CM) side (left) and cardiac fibroblast (CF) side (right) for samples with spontaneously beating cells (d, top) and for samples stimulated from the CM side (d, middle) and the CF side (d, bottom) sequentially.

Vimentin (CF marker) and Cardiac Troponin-I (CM marker) staining of the MCDs. Immunostaining data confirmed that there were no CMs on the CF side of the pattern and that CFs were able to proliferate towards the protein side and complete the structure as expected. Therefore, our self-forming micropatterning approach was successfully implemented.

Various ion channels contribute to the excitability of CMs.[27] However, CFs do not have the same type, distribution and density of such channels, and thus cannot fire APs upon an input.[28, 35, 36] For example, unless genetically modified,[28, 29] CFs lack most of the voltage sensitive $K^+$ channels,[30] which are crucial for excitability.[27] Stimulations from CM side are received through these voltage sensitive ion channels and APs are propagated through gap junctions (Figure 3b). Figure 3c shows, nuclei (blue) of both CMs and CFs, striated CMs (green), and the gap junctions (red) between CM – CM, CM – CF and CF – CF, which are crucial for intercellular ion transportation.

Figure 3d, shows membrane potential measurements on MCDs using the MEAs. Throughout these measurements we monitored the MCD to confirm the presence of healthy, beating cardiac cells on the excitable half of the MCD (Supplementary Movie 2). We measured the spontaneous membrane potentials of the cells of the MCD to be lower than that of the cells in the unconfined, patterned co-culture for both CMs and CFs (Figure 3d, top). However, this voltage was sufficient to illustrate unidirectional signal propagation through the MCD. In future studies, to improve electrical activity of the micropatterned cells, the protein pattern could be modified to provide an anisotropic alignment to seeded cells.[37] Similar to previous measurements with unconfined, micropatterned co-culture, we stimulated and measured the MCD from both CM and CF ends. In the forward direction, upon electrical input using the MEAs, the CMs were excited and the signals propagating through gap junctions were measured from the CF side (Figure 3d, middle). In the reverse direction, the CFs could not be excited upon the same magnitude of electrical stimulation since they lack the proper ion channels on their membranes, thus the signal could not be amplified and propagated, and there was no detectable output signal (Figure 3d, bottom). These results showed that the MCD successfully operates as a diode by propagating the applied signal unidirectionally, and that the cells preserved their transport properties even under confinement.

A solid-state diode is an electrical circuit component that shows nonlinear current-voltage characteristics allowing current to flow in only one direction. An ideal diode is a switch that opens when current is flowing in a certain direction and is closed otherwise. This basic binary function makes diodes a key component of logic operations including, but not limited to, 'AND' and 'OR' gates. Recently, iontronics, control of ions and ionic flows for information processing, has started to gain greater interest as an alternative to solid-state electronics.[38, 39] This technology attempts to mimic the phenomena in nature where ion transport is precisely controlled through cell membranes via ion selective channels in order to perform cellular functions. While solid state electronics use electron and hole transport through semiconductor junctions, iontronics use ion transport through ion-selective membranes. This basic difference in working principle, as well as high level of degeneracy, gives iontronics the capability to better mimic biological systems.[40] Contrary to current iontronics approaches, in this study, we control the organization of excitable and non-excitable cells and manipulate cellular connections to directly control ionic transport for information processing instead of mimicking cell membrane behavior using synthetic ionic membranes. While other iontronic approaches attempt to control ion flow in a manner analogous to metal-oxide-semiconductor field-effect-transistors, our diode approach is more similar to pn junctions of bipolar junction transistors. Both of these complementary approaches based on ionic currents can be used in aqueous media conditions and may be easily integrated to constitute stronger computational capacity for biological applications. Furthermore, the conduction velocity of the CMs, which was calculated to be 30.0 ± 1.3 cm/s, was comparable with that of ionic devices (~ 100 cm/s) supporting the possibility of combining these two types of devices.[40]

A biological diode similar to the one we designed and fabricated, is a starting point for designing more complex cell-based electrical components (e.g., transistors) and eventually biological logic gates and processors. Such cell-based ionic circuits and circuit networks can be electrically coupled to traditional circuits or sensors and offer a new approach for directly and precisely linking communication systems in a living system, such as neural implants or interfaces of muscle cells with electrical devices. Furthermore, untraditional sensing or transducing approaches can be used in combination with our cell-based electronics. For example, organic electrochemical transistors are recently introduced as bioelectrical sensors where ionic signals are transduced into electrical signals.[41, 42] Muscle cell-based electrical components that naturally function via ionic currents could provide a novel insight into signal communication and processing in living systems when combined with such emerging technologies.

Although there are some studies recapitulating neural cell networks as circuit elements,[43, 44] but none that uses muscle cells. While individual neurons can propagate signals unidirectionally, sophisticated techniques are required to control the orientation of emitting and receiving neuron networks.[43] In contrast, the MCD approach described here does not require such techniques and, more importantly, promises absolute unidirectional signal transduction. Feinerman et al. showed the nonlinear response of patterned neural cells to external inputs.[44] This nonlinearity was caused by a threshold which all excitable cells require to fire. However, the unidirectionality in electrical signal transduction that we achieved using muscle cells was not as successful with neural cells. The neural diode had an 8% error in the reverse direction, whereas in our design it is biologically impossible to pass signals in the reverse direction (i.e., CF to CM). Furthermore, the pacing ability of the CMs demonstrated here allows us to modulate the frequency of their electrical activity and therefore pass information embedded in the electrical signal.

MCD designed and fabricated in this study is a proof of concept that live cells can be organized to process logic operations, and eventually, as electromechanical processors. Thus, muscle cell networks that can replicate a diode function are not only a novel platform for studying interactions between muscle cells, but more generally, a new approach for bioelectrical and biomechanical interfaces and biocomputing. Such networks can be used safely in clinical settings as a human-electronics interface, since it does not require any genetic manipulation of the cells in the device, and avoids the addition of bioactive and/or chemical agents to achieve the desired output. The implementation and investigation of MCDs will impact the fundamental understanding of cell-cell and cell-environment communication in muscle cell networks. Utilizing this knowledge, MCDs and more sophisticated living logic devices will transform how bioelectrical and biomechanical interfaces are engineered.

Heart muscle cells can be electrically or mechanically stimulated through their voltage sensitive or mechanosensitive ion channels, respectively. Here we show that, these electrically and mechanically responsive heart muscle cells are ideal candidates for cell-based information processing, since, when organized properly, they could be used to control the information flow. Such cell-based circuit components can pave the way for cell-based electromechanical circuits and circuit networks that can couple to traditional circuits or sensors. They therefore offer a new approach for directly linking the communication circuitry in a living system, e.g., neural circuitry or muscle cell contraction signaling, with electrical or mechanical devices. Furthermore, these 'living circuitries' can be directly used as control units for other biomedical engineering applications such as bioactuators or biosensors. Cell networks that can replicate diode and logic gate functions with muscle cells is a novel platform for studying muscle cell interactions and, more generally, a new approach for bioelectrical and biomechanical interfaces and biocomputing.

## Experimental Section

*Fabrication of PDMS Stencils and Sheets*: In order to create micropatterned surfaces, SU-8 2075 (MicroChem Corp.) photoresist was spin coated (1000 rpm, 300 rpm/s, 30 s) to obtain a thickness of 200 ± 20 μm on a silicon (Si) wafer (Universiry Wafer), following manufacturer's instructions. PDMS (Ellsworth Adhesives) base and curing agent were mixed in 5:1 ratio, degassed, spin coated on the silicon wafers (750 rpm, 100 rpm/s, 30 s) and cured at 70°C for 30 minutes.

*Cell Isolation and Culture:* Micropatterned substrates were seeded with neonatal rat ventricular cardiac cells isolated according to a previously established protocol[45] and following regulations of University of Notre Dame's Institutional Animal Care and Use Committee. The culture was maintained under standard cell culture conditions in Dulbecco's Modified Eagle Medium (DMEM, Hyclone) supplemented with fetal bovine serum (FBS, 10%, Hyclone) and penicillin-streptomycin (P/S, 1%, Corning). Endogenous fibronectin was removed from the FBS using gelatin sepharose 4B (GE Healthcare).

*Fabrication of the MCD:* Fibronectin (50 μg/mL, Sigma-Aldrich) / Alexa-488 or Alexa-647 tagged fibrinogen (50 μg/mL, Molecular Probes) solution was added on top of the stencil and was incubated at 37°C for 30 minutes. Following a phosphate buffered saline (PBS, Corning) wash, stencils were removed. Then MEA surfaces were coated with Pluronic F127 (1% solution in water, Sigma-Aldrich), for 1 hour. CM enriched, CF containing cell suspension was seeded at a density of $0.5 \times 10^6$ cells/ml and incubated overnight and the PDMS sheet was peeled off. In 4-5 days the CFs proliferated to fill the pattern.

*$Ca^{2+}$ Indicator Loading:* Co-culture was loaded with Fluo-4 acetoxymethyl ester (Molecular Probes), which exhibits increase in fluorescence intensity upon binding to $Ca^{2+}$, following manufacturer's instructions.

*Electrical Signal Measurements and Stimulations:* Electrical field potential measurements were performed using the MEA-2100 system (Multichannel Systems) with a sampling rate of 2.5 kHz. Cells were stimulated with ±400 mV, 1 ms biphasic pulses of various frequencies (i.e., 1Hz, 2 Hz, 3Hz). Biphasic pulses were achieved by using two electrodes simultaneously for stimulations.

*Data Acqusition and Plotting:* Data sets from electrical measurements were exported and plotted using MATLAB. All data sets (spontaneous activity and response to stimulations) were collected from both the CM and CF sides of the culture simultaneously. For the spontaneous activity measurements, each individual AP was detected by a 40 μV treshold from the CM side. For the stimulation measurements, the signals collected were plotted using the stimulation instant (precisely defined by the input signal) as t = 1 μs for each individual stimulation. For all cases these signals were plotted using raw data (gray curves) and then averaged (red and green curves). The distance between two electrodes of the MEA was divided by the time the AP required to propagate from one electrode to another in conduction velocity calculations. This time difference was calculated by comparing the times measured from the these two electrodes when the maximum voltage occurs.

## Acknowledgements

This study is supported by NSF Grant No: 1530884. Photolithography steps were conducted in the cleanroom facility of University of Notre Dame Nanofabrication Facility. Confocal imaging was performed at Notre Dame Integrated Imaging Facility.

# Supporting Information

Most electrically excitable cell models start with an equation of the form $\frac{dV}{dt} = \frac{-I_{ion}}{C}$ where $V$ is the membrane potential difference, $I_{ion}$ is the sum of all ion currents that move through the ion channels and $C$ is the membrane capacitance. The electrical activity of a cell is more complicated than what this equation reveals, since $I_{ion}$ is the sum of multiple transmembrane currents (e.g. $I_{ion} = I_{Na+} + I_{Ca2+} + I_{K+} + I_{Cl-} + \cdots$) each of which may couple back to one another, $C$, $V$ and other intracellular processes. In the simplest case, currents are taken proportional to the potential difference from a fixed reference (e.g., $I_{Na+} = g_{Na+}(V - E_{Na+})$) with no time dependence in membrane conductivity $g$ and rest potential $E$.

For the computational model, we adapt the model of Karma[14] to include an arrangement of excitable and non-excitable cells. We coarse grain individual cells, couple nearest neighbors linearly, and assign parameter values spatially according to our excitable-non-excitable diode pattern.

$$\frac{\partial V(x,t)}{\partial t} = \nabla \cdot (D(x)\nabla V(x,t)) + w_v(x)\left[-V(x,t) + \left(A - \frac{n(x,t)}{v_s}\right)^m \left(1 - tanh(V(x,t) - v_h)\frac{V^2(x,t)}{2}\right)\right] \quad (1)$$

$$\frac{\partial n(x,t)}{\partial t} = w_n(x)\left[\theta\big(V(x,t) - v_n\mathcal{R}(n)\big) - (1 - \theta(V(x,t) - v_n))n(x,t)\right] \quad (2)$$

Here $V(t)$ represents the change in membrane potential from its resting point, whereas, $n(t)$ is an "effective variable" representing the cumulative change in the conductivity of ion gates. The change in the membrane potential of a cell at time t, at position x depends on the difference of its membrane potential with that of the neighboring cells (the first term on the right hand side of Eqn.1) and the "excitability state" of the cell characterized by both $V(x,t)$ and $n(x,t)$ (the second term on the right hand side of Eqn.1); $\mathcal{R}(n) = \frac{1-(1-e^{-r})n(x,t)}{1-e^{-r}}$ is the "restitution function", $\theta$ is the unit step function, and $w_v$ and $w_n$ determine the time scale of excitation of $V$ and $n$, respectively. We took the interaction coefficient for the CM – CM, CM – CF and CF – CF junctions to be equal, with $D = 0.05$ cm$^2$/ms. Modifying this assumption does not make a significant difference in the qualitative characteristics of the propagating signal. We introduce a position dependence to the parameter $w_v$ since its value depends on whether the cell at $x$ is a CF or a CM. Since the CFs are non-excitable, we took $w_v(x)$ to drop discontinuously from 0.17 ms$^{-1}$ to 0 at the interface between the two cell groups. All remaining parameter values were chosen consistently with the original Karma model of a single excitable cell.[14] We used A=1.54, m=10, $w_n = 0.001$, r=1, $v_s = 0.64$, $v_h = 3$, $v_n =1$ units, consistent with the basic original model. We integrated the equations with dt=0.1 ms and dx=10$^{-3}$ cm using the Euler method with Mathemetica (Wolfram Research).

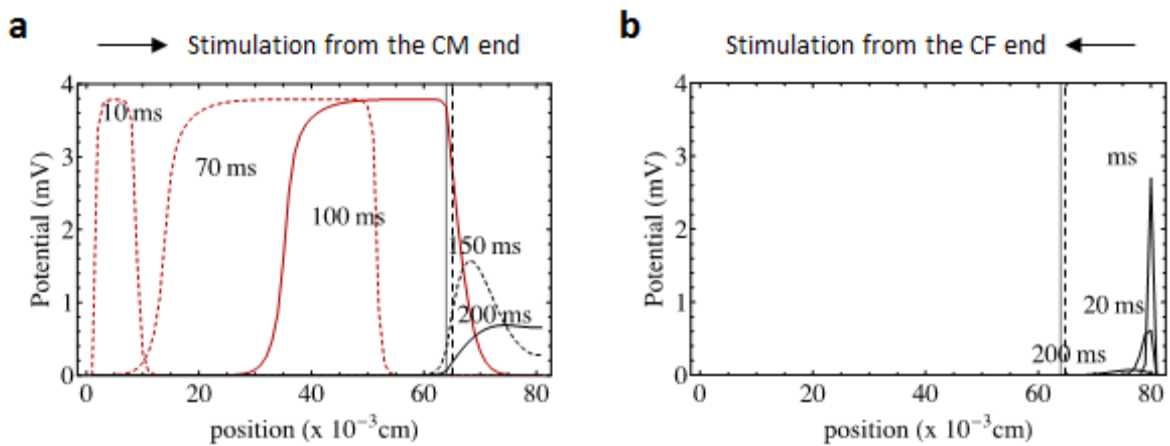

**Figure S1.** Computer simulation of electrical stimulus (3 mV for 0.1 ms) propagation within a micropatterned excitable CM (red curves) / non-excitable, CF (black curves) co-culture in forward, from the CM end (a) and in reverse, from the CF end (b) directions. The dashed vertical line marks the CM – CF boundary. (a) shows a readout of the signal on the right side of the cell domain, when the signal originates from the left side. (b) shows no readout of the signal on the left side when the signal originates from the right side.

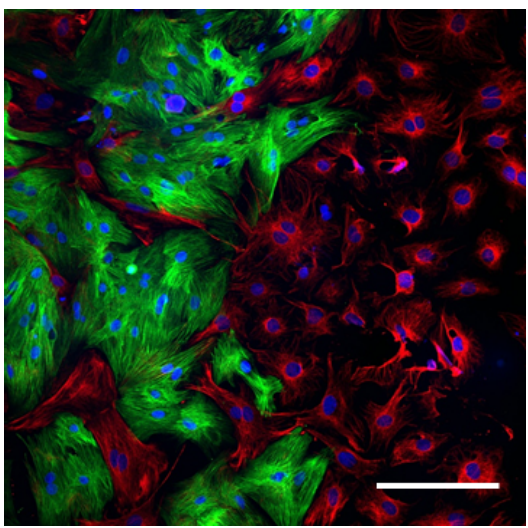

**Figure S2.** Fluorescence image of Troponin-I (green) and Connexin 43 (red) immunostaining of the patterned co-culture counter stained for the cell nuclei (blue) (c, scale bar: 100 μm) for the unconfined, micropatterned sample (scale bar: 100 μm).

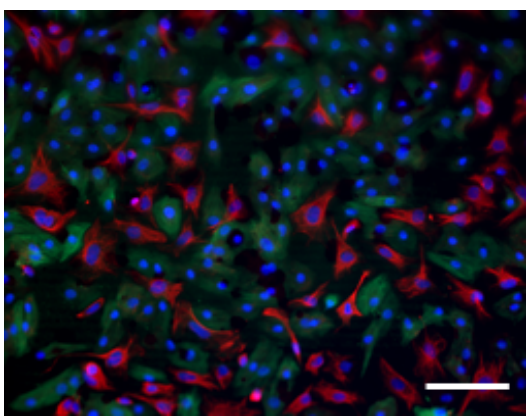

**Figure S3.** Fluorescence image of Troponin-I (green) and Connexin 43 (red) immunostaining of the CM enriched cell suspension counter stained for the cell nuclei (blue). Cells were seeded and fixed on Day 1 to determine the cell type ratios (scale bar: 100 μm).

**Supplementary Movie 1.** $Ca^{2+}$ video captured during spontaneous contractions of CM (left bottom half) and CF (right upper half) co-culture.

**Supplementary Movie 2.** Phase contrast video of the MCD captured during spontaneous electrical activity measurement on an MEA.

**Extended Experimental Section**

*Fabrication of PDMS stencils:* In order to create micropatterned surfaces, SU-8 2075 (MicroChem Corp.) photoresist was spin coated (1000 rpm, 300 rpm/s, 30 s) to obtain a thickness of 200 ± 20 μm on a silicon (Si) wafer (Universiry Wafer). The Si wafer was soft baked for 10 minutes at 65°C and 45 minutes at 95°C followed by UV exposure through a transperency mask (Advanced Reproductions) using a mask aligner (Karl Suss MJB-3), and then was developed using SU8 developer (MicroChem Corp.). PDMS (Ellsworth Adhesives) base and curing agent were mixed in 5:1 ratio, degassed, spin coated on the silicon wafers (750 rpm, 100 rpm/s, 30 s) and cured at 70°C for 30 minutes. Finally, the thin film PDMS was carefully peeled off to obtain the elastomeric stencils with the desired patterns and sterilized under UV prior to cell culture use.

*Protein Patterning:* Stencils having rectangular patterns (500 x 1000 μm) were aligned and placed on substrates containing embedded MEAs (Multichannel Systems). Fibronectin (50 μg/mL, Sigma-Aldrich) / Alexa-488 or Alexa-647 tagged fibrinogen (50 μg/mL, Molecular Probes) solution was added on top of the stencil and was incubated at 37°C for 30 minutes. Following a phosphate buffered saline (PBS, Corning) wash, stencils were removed to obtain fibronectin patterns. In order to prevent non-specific cell attachment to the outside of the protein pattern, the MEA surfaces were coated with an antifouling agent, Pluronic F127 (1% solution in water, Sigma-Aldrich), for 1 hour. The substrates were then washed several times with PBS to remove the remaining Pluronic F127. Prior to cell seeding, thin PDMS sheets prepared by spin coating, as described above for PDMS stencils, were placed on the protein patterns in a way that covers half of the pattern to prevent initial cell attachment.

*Cell isolation, culture and characterization:* Micropatterned substrates were seeded with neonatal rat ventricular cardiac cells isolated according to a previously established protocol [33] and following regulations of University of Notre Dame's Institutional Animal Care and Use Committee. Briefly, the hearts were excised from 2-day old neonatal Sprague-Dawley rat pups, diced into small parts, incubated overnight in 0.05% (w/v) Trypsin (Gibco) in Hank's Balanced Salt Solution (HBSS,

Gibco) followed by 0.1% collagenase type-2 treatment. All the isolated cells were placed on tissue culture plates. Since CMs require more time to attach to tissue culture substrate, the first cells attached were CFs. Other cells that are present in the heart wall tissue (i.e., endothelial cells) were mostly eliminated due to the specific media used. The unattached cells at the end of the 1.5 hour pre-plating, mostly CMs, were collected and seeded onto MEA substrates. The culture was maintained under standard cell culture conditions in Dulbecco's Modified Eagle Medium (DMEM, Hyclone) supplemented with fetal bovine serum (FBS, 10%, Hyclone) and penicillin-streptomycin (P/S, 1%, Corning). Endogenous fibronectin was removed from the FBS using gelatin sepharose 4B (GE Healthcare).

*Fabrication of the MCD:* CM enriched, CF containing cell suspension was seeded at a density of $0.5 \times 10^6$ cells/ml onto the half-covered fibronectin patterns and incubated overnight. The MEAs were then washed with PBS to remove any unattached and/or weakly attached cells. Once all the unattached cells were removed from the culture, the PDMS sheet was peeled off. As the CMs are non-dividing cells, only CFs proliferated to the fibronectin micropattern previously covered by the PDMS sheet. In 4-5 days, the CFs proliferated to fill the pattern and a configuration where half the pattern is excitable and the other half is non-excitable was achieved.

*$Ca^{2+}$ Indicator Loading:* Co-culture was loaded with Fluo-4 acetoxymethyl ester (Molecular Probes), which exhibits increase in fluorescence intensity upon binding to $Ca^{2+}$, following manufacturer's instructions. Briefly, the co-culture was incubated in Tyrode's salt solution (Sigma Aldrich) loaded with 3µM Fluo-4 acetoxymethyl ester and 0.02% Pluronic F127 (Life Technologies) and incubated at 37 °C for 30 min. Then washed with PBS and kept in normal culture medium. $Ca^{2+}$ fluxing during spontaneous contractions of the micropatterned CM – CF co-culture was captured using high-speed florescence microscope imaging (Axio Observer.Z1, Carl Zeiss).

*Electrical Signal Measurements and Stimulations:* MEAs in this study consist of poly-3,4-ethylenedioxythiophene – carbon nanotube (PEDOT – CNT) electrodes with electrode spacing of 200 μm and electrode diameter of 30 μm. Electrical field potential measurements from these electrodes were performed using the MEA-2100 system (Multichannel Systems) with a sampling rate of 2.5 kHz. Briefly, the MEAs were placed onto the head stage to read or write the electrical signals through the contact pads of the MEAs. The signals read by head stage pins were transferred to a PC using an interface board. The temperature was kept constant at 37°C throughout the experiments by a temperature controller unit (TC02, Multichannel Systems). Cells were stimulated with ±400 mV, 1 ms biphasic pulses of various frequencies (i.e., 1Hz, 2 Hz, 3Hz). The MEAs used in this study allow simultaneous stimulation and recording of electrical signals for up to 60 channels, with the capability of assigning recording or stimulation functions to individual channels. Biphasic pulses were achieved by using two electrodes simultaneously for stimulations.

*Data Acqusition and Plotting:* Data sets from electrical measurements were exported and plotted using MATLAB. All data sets (spontaneous activity and response to stimulations) were collected from both the CM and CF sides of the culture simultaneously. For the spontaneous activity measurements, each individual AP was detected by a 40 μV treshold from the CM side. These APs were then plotted for both the CM and CF sides since the measurements are simultaneous (Figure 4a,5c). For the stimulation measurements, the signals collected were plotted using the stimulation instant (precisely defined by the input signal) as t = 1 μs for each individual stimulation (Figure 4b,c,5d,e). For all cases these signals were plotted using raw data (Figure 4,5c-e, gray curves) and then averaged (Figure 4,5c-e, red and green curves). The distance between two electrodes of the MEA was divided by the time the AP required to propagate from one electrode to another in CV calculations. This time difference was calculated by comparing the times measured from the these two electrodes when the maximum voltage occurs.

*Immunostaining:* Cells were washed with PBS and fixed by incubating in paraformaldehyde (4%, Electron Microscopy Sciences) for 15 minutes at room temperature, and then washed with PBS. Cells were then permeabilized in Triton X (1%, Sigma-Aldrich) for 15 minutes and then washed with PBS. Cells were blocked with goat serum (10%, Sigma-Aldrich) for two hours. After blocking, cells were incubated with mouse monoclonal cardiac Troponin-I (Abcam) primary antibody diluted (1:100) in goat serum at 4°C overnight. The next day, cells were washed with PBS and then incubated with Alexa Fluor 488 (Life-Technologies) secondary antibody diluted (1:200) in goat serum at 4°C for 6 hours. Following the secondary antibody incubation, cells were washed with PBS. The steps were repeated for rabbit monoclonal cardiac Vimentin (Abcam) or rabbit monoclonal cardiac Connexin 43 (Abcam) primary antibodies and Alexa Fluor 594 (Life-Technologies) secondary antibody. After a second staining, cells were incubated with nuclear stain DAPI (1:1000 DAPI:PBS, Sigma Aldrich) and then washed with PBS until no background was seen. Imaging was performed using confocal microscopy (Nikon C2+).